\documentstyle[twocolumn,prl,aps,epsfig]{revtex}
\begin{document}

\title{\bf
A New Feature in Some Quasi-discontinuous Systems \footnote{
Supported by the National Natural Science Foundation of China
under grant No. 19975039, and the Foundation of Jiangsu Provincial
Education Committee under the Grant No. 98kjb140006.
}
}
\author{
WU Shun-Guang(\hspace{2cm}) $^{1,2}$ \qquad
 HE Da-Ren(\hspace{2cm}) $^{1}$\\
}
\address{
$^1$ Complexity Science Center, Yangzhou University,
Yangzhou 225002\\
$^2$ Institute of Low Energy Nuclear Physics, Beijing Normal University, 
Beijing 100875 \\
\quad\\
(Received 20 October, 1999)\\
}
\maketitle

\vspace{1cm}
{\small Many systems can display a very short, rapid changing stage
(quasi-discontinuous region) inside a relatively very long and slowly
 changing process. A quantitative definition for the "quasi-discontinuity"
 in these systems has been introduced. We have shown by a simplified model
 that extra-large Feigenbaum constants can be found inside some period-doubling cascades
  due to the quasi-discontinuity. As an example, this phenomenon has also
  been observed in Rose-Hindmash model describing neuron activities.\\
}

Recently, there has been considerable interest in piece-wise smooth systems
(PWSSs). Such models usually describe systems displaying sudden,
discontinuous changes, or jumping transitions after a long, gradually
varying process. These systems may show
 some behaviors apparently different from those of the
 everywhere - differentiable systems (EDSs) [1-5].
In fact, the sudden changes in the above processes also need time. Therefore,
such a process can be everywhere smooth if one describes it with a high
enough resolution. Usually, in the largest part of the process, a quantity
 changes very slowly. It has a drastic changing only in one or several
 very small stages. We suggest to call the stage as a "quasi-discontinuous
 region (QDR)" and shall define a "quasi-discontinuity (QD)" inside it
 quantitatively. A system that can display QDR in its processes may
  be called a "quasi-discontinuous system (QDS)". Obviously, QDS is a
  much wider conception than PWSS and may serve as an intermediate between
   EDS and PWSS. 

In order to show our basic idea and the first characteristic
of QDS, we have
 constructed a model map as shown in Fig.1. The map reads:
\begin{equation}
f(x)=
   \left\{
       \begin{array}{ll}
	 f_{1}(x)=k_1(x-x_1)+y_1        & x\in [0,x_1),\\
	 f_{2}(x)=A(x-x_0)^2+y_0        & x\in [x_1,x_3),\\
	 f_{3}(x)=\sqrt{r^2-(x-o_x)^2}+o_y      & x\in[x_3,x_4),\\
	 f_{4}(x)=k_2(x-x_4)+y_4             & x\in[x_4,x_5),\\
	 f_{5}(x)=k_3(x-1)             & x\in[x_5,1].
       \end{array}
   \right.
\label{eq31}
\end{equation}
As can be seen in Fig.1, the slope of $f_1$ branch is a unit. It is
 simulating the slowly changing part of the process.  Branch $f_4$
  is a linear line with a very large negative slope $k_2$. Branch $f_3$
   is a small part of a circle introduced for a smooth connection of
   $f_2$ and $f_4$. The center of the circle locates at ($o_x,o_y$),
    and its radius is $r$. Branches $f_3$ and $f_4$ can simulate the small
    drastic changing part. In Eqs. (1), $A$ is chosen as the control
    parameter. It is obvious that the fixed point at $P_2$ 
    undergoes period-doubling bifurcation when $A$ changes inside a
    certain parameter range. $(x_j,y_j)$ denote the coordinates of points $P_j$
     (j=0,...,5), respectively. They are determined by the conditions of smooth
     connections between neighboring branches. For certain function forms
      of $f_2$ and $f_4$, the circle of $f_3$ still may be very large or
      small. We define another parameter $\alpha$ to fix it. Therefore,
      $o_x$, $o_y$, $r$, $k$, and $(x_j,y_j)$ are all functions of $A$ and
      $\alpha$. Their explicit forms will not be shown in this short letter. The parameter ranges chosen for this
 study are $A \in [8.0,9.0]$ and $\alpha \in [0.99995,0.999997]$.

Now we will define QDR and QD in this model.
According to the geometrical properties of Eqs. (1), one can obtain the following
 conclusions. When
 $\alpha=1$, $r=0$, the first order derivative of the map function
 is discontinuous at $x_3=x_4$. The second order derivative shows a
 singularity, that is, an infinitely large value here. When $\alpha \in (0, 1)$,
 the branch $f_3$ has a finite length. The first order derivative of
 the map function is continuous at both $x_3$ and $x_4$. The second
 order derivative value between them is finite but trends toward infinite
  when $ \alpha \rightarrow 1$. In this case, the maximum value of the
  second order derivative of the map function between $x_3$ and $x_4$
  may be used to describe the "quasi-discontinuity (QD)". So we
  shall  define QD as
\begin{equation}
\kappa=max{|\frac{d^2f}{dx^2}|_{x_0}, x_0\in{[x_3,x_4]}},
\end{equation}
and define QDR as
\begin{equation}
\Delta=|x(2)-x(1)|,
\end{equation}
where $x(2)$ and $x(1)$ are between $x_3$ and $x_4$, and satisfy
$$
|\frac{df}{dx}|_{x(1)}=|\frac{df}{dx}|_{x(2)}=
\frac{1}{\sqrt{2}}\max{|\frac{df}{dx}|}.
$$

Acording to the definations (2) and (3), the QDR and QD for Eqs. (1)
can be expressed as
\begin{equation}
x(1)=o_x -\frac{k_2 r}{\sqrt{2+k_2^2}}, \quad \quad x(2)=x_4,
\end{equation}
and
\begin{equation}
\kappa=\frac{-r^2}{[r^2-(x_4-o_x)^2]^{3/2}},
\end{equation}
respectively.

When $\alpha=1$ it is
reasonable to observe one of the typical behaviors of PWSS.
That is the interruption of a period - doubling bifurcation cascades by
 a type V intermittency [2,3]. 

When $\alpha$ is smaller than, but close to 1, there is a QDR
between $x_3$ and $x_4$
instead of the non-differentiable point. The mapping is everywhere smooth,
so the period-doubling bifurcation cascade should continue to the end.
However, there is a drastic transition of the mapping function slope
in a very small QDR that makes all further bifurcation points compressed
 into a relatively much shorter parameter
distance. The Feigenbaum constants $\delta_i$ (i=3,4,5 or even more),
influenced by
the compression, should show some extraordinary values. That is exactly
what we have
observed. Table 1 shows the data about three cascades. In the table, $n$ indicate
 the sequence number of doubling, $\delta_n(i)$ $(i=1-3)$ are the Feigenbaum
 constants of the cascade $i$. The 
parameter values $\alpha(i)$ and the
maximum value of the QD, $\kappa(i)$, for each cascade are 
indicated in the caption. In the table $\delta_{n_0}$ data are obtained from
 Ref. [6]. They are listed here for 
a comparison with the corresponding ones obtained in a
typical everywhere smooth
situation. As can be seen in table I, when $\kappa(i)$ is large,
a lot of Feigenbaum
constants, $\delta_3$, $\delta_4$, $\delta_5$, $\delta_6$
and $\delta_7$ are
extraordinary. The further constants may be considered as ordinary,
but they converge
to the universal Feigenbaum number very slowly. When $\kappa(i)$ is smaller, only 
$\delta_3$, $\delta_4$ and $\delta_5$ are apparently extraordinary. The further 
constants converge much faster. When $\kappa(i)$ is very small, the whole Feigenbaum 
constant sequence is very close to the standard $\delta_{n_0}$ data. That may
 indicate a smooth transition from QDS to a EDS.
Also, from these data one can believe that
the extraordinary Feigenbaum constants in the period-doubling cascades
are induced
by QD of the system. Based on this understanding we suggest the use of
 the common
extraordinary Feigenbaum constant $\delta_{j} (j=3,4)$ to signify this
 phenomenon.
The relationship between $\kappa$, the QD, and the symbol of the phenomenon 
$\delta_{j}$ ($j=3,4$), have been computed. Figure 2 shows the result of
 function
$\kappa-\delta_{j}$(
Although $\kappa$ is dependent on both $\alpha$ and $A$,
our numerical results demonstrate that $\kappa$ is not sensitive
to the parameter $A$ at a given $\alpha$.
Therefore, it is possible to choose the maxmum $\kappa$ to represent QD of
a whole diagram. For example,
for the bifurcation points $A_n(n=0,1,2, ..., 12)$, indicated in the second
column of Table 1, the corresponding
$\kappa$ are $220.9 \times 10^6$, $208.2 \times 10^6$, $205.6 \times 10^6$, ...,
$205.0 \times 10^6$, respectively. So we choose
$220.9 \times 10^6$ as the representative $\kappa$ of the bifurcation diagram).
 One can see that $\delta_{4}$ increases, but $\delta_{3}$
decreases when $\kappa$ becomes larger and larger.

It is important to find examples of this kind of interesting phenomenon in 
practical systems. We have done such a study in Rose-Hindmarsh (R-H) model.
The model, which describes neuronal bursting [7], can be expressed by 
 \begin{equation}
   \left\{
       \begin{array}{l}
	  \frac{dx}{dt}=y-ax^3+bx^2+I-z,\\
	  \frac{dy}{dt}=c-dx^2-y,\\
	  \frac{dz}{dt}=r[s(x-x^*)-z],
       \end{array}
   \right.
\end{equation}
where $x$ is the electrical potential of the biology membrane, $y$ is
 the recovering variable, $z$ is the adjusting current, $a,b,c,d,s$
 and $x^*$ are constants, $r$ and $I$ are chosen as the control
 parameters. We shall take $a=1,b=3,c=1,d=5,x^*=1.6,s=4$ for this study.
 Fig.3 shows the ${\rm Poincar\grave{e}}$ map of a strange attractor
 observed when $I=2.9, r=0.00433$
(Here the ${\rm Poincar\grave{e}}$ section is defined as the coordinate
value of z axis at the maximum in x direction of the trajectory. We have also
tested some different definitions of ${\rm Poincar\grave{e}}$ section,
the results have
shown that all of them are qualitatively the same as each other).
 It is clear that the
 iterations in the region $[z_1,z_2]$ change very rapidly. Therefore,
 we call this region as a QDR.

 Table 2 shows the critical bifurcation parameter values and the corresponding
   Feigenbaum constants for a period-doubling bifurcation cascade.
   One can see that $\delta_1$ and $\delta_2$ are
     larger than ordinary values. As our computation has confirmed, that
      means an interruption of the cascade by a collision of the periodic
       orbit with the QDR when the first time period-doubling finished.

For a comparison with the function $\kappa-\delta_{i}$ shown in Fig. 2,
 we have computed the bifurcation diagrams with $I=2.8,2.9,3.0,3.1,
 3.2,3.3,3.4$, and $r \in [0.1\times 10^{-2},4\times 10^{-2}]$.
 The results are shown in Fig. 4 (Here, we also choose the
maxmum $\kappa$ to represent the quasi-discontinuouty of one 
bifurcation diagram). They are in a qualitative agreement
with those in Fig. 2.

In conclusion, we have found some extraordinary Feigenbaum constants in some
period-doubling bifurcation cascades in a constructive and a practical system.
The mechanism of the phenomenon is that a periodic orbit near a critical point of bifurcation crosses a QDR in the system.
This understanding may be important for 
the experimental scientists because very often they can measure
only the first several
Feigenbaum constants in a real experiment. After observing strange Feigenbaum 
constants, they can verify if their system is a QDS with the knowledge in this 
discussion. 
Moreover, our results also demonstrate that between typical PWSSs and EDSs 
there can be a type of transitive systems.

\begin{table}
  \caption{ The data for period-doubling bifurcation cascades in
  Eqs. (1) with
  $\alpha=0.9999970$ and $\kappa_{max}=220.9 \times 10^6$ for cascade (1); 
$\alpha=0.9999196$  and $\kappa_{max}=8.2 \times 10^6$ for cascade (2); 
$\alpha=0.9987537$ and $\kappa_{max}=0.5 \times 10^6$ for cascade (3).
The data are obtained by numerical method with a quadruple precision.
}
  \begin{center}
     \begin{tabular}{c|c|c|c|c}
	{\sl n} & {\sl $\delta_n(1)$} & {\sl $\delta_n(2)$} & {\sl $\delta_n(3)$} & {\sl $\delta_{n_0}$}\\
	\hline
  1&  4.626168416 & 4.626168416 & 4.626168416 & 4.744309468 \\
  2&  4.638635250 & 4.638635250 & 5.965311262 & 4.674447827 \\
  3&  8.823236594 & 9.982064249 & 7.746315906 & 4.670792250\\
  4& 16.553602405 &11.112580986 & 4.756159889 & 4.669461648\\
  5&  1.619920263 & 3.966760226 & 4.705758197 & 4.669265809 \\ 
  6&  4.147211067 & 4.411839476 & 4.674983020 & 4.669214270\\  
  7&  5.173386810 & 4.526007508 & 4.670687838 & 4.669204451 \\  
  8&  4.765263421 & 4.709611842 & 4.669607200 & 4.669202201 \\  
  9&  4.757647408 & 4.671017586 & 4.669986034 & 4.669201737 \\  
 10&  4.687625667 & 4.677922775 & 4.669375410 & 4.669201636 \\  
 11&  4.701929236 & 4.703838840 & 4.699239861 & 4.669201614\\  
     \end{tabular}
  \end{center}
\end{table}

\begin{table}
  \caption{The data of the bifurcation critical parameter values and the
   corresponding Feigenbaum constants observed in system (6) with $I=2.8$.}
  \begin{center}
     \begin{tabular}{c|c|c}
	{\sl n} & {\sl $r_n\times 10^{-2}$} & {\sl $\delta_n$}\\
	\hline
	  0 &       3.52000 &  \\      
	  1 &       1.55550 &      19.84344\\
	  2 &       1.45650 &       6.14905 \\
	  3 &       1.44040 &       5.19350 \\
	  4 &       1.43730 &       4.76981 \\
	  5 &       1.43665 & \\    
     \end{tabular}
  \end{center}
\end{table}

\begin{figure}
\label{ff1}
\centerline{\epsfig{figure=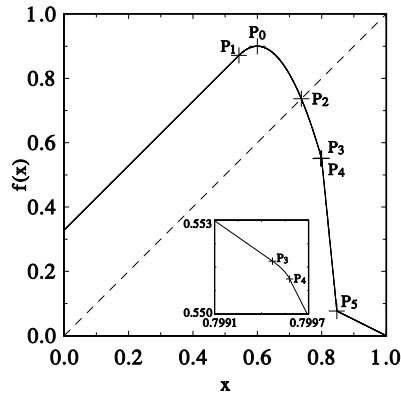,width=8cm,angle=0}}
\caption{Function of Eqs. (1) when $A=8.75407, \alpha=0.998$.
The inset is the enlargement area near $P_3$ and $P_4$. }
\end{figure}

\begin{figure}
\label{ff2}
\centerline{\epsfig{figure=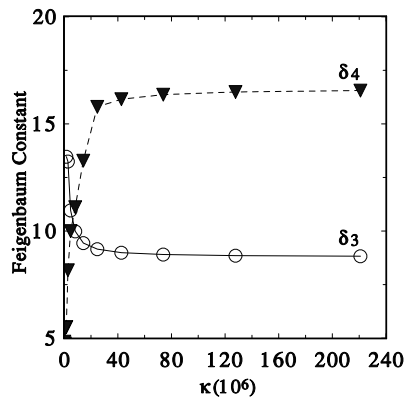,width=8cm,angle=0}}
\caption{ The function $\kappa-\delta_{i}$ for Eqs. (1)
with the parameter value $\alpha \in [0.99966,0.999997]$.}
\end{figure}

\begin{figure}
\label{ff3}
\centerline{\epsfig{figure=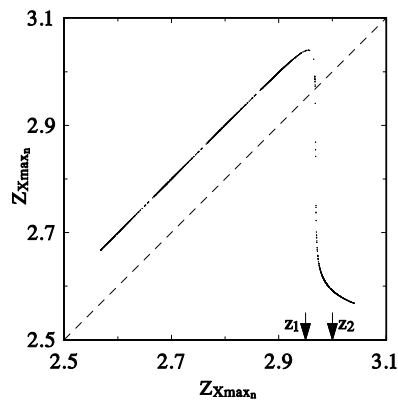,width=8cm,angle=0}}
\caption{
The ${\rm Poincar\grave{e}}$ map of R-H model. Here the range
$[z_1, z_2]$ schematically indicates the QDR, the real QDR calculated
via Eq.(3) is too small to be clearly expressed.
}
\end{figure}

\begin{figure}
\label{ff4}
\centerline{\epsfig{figure=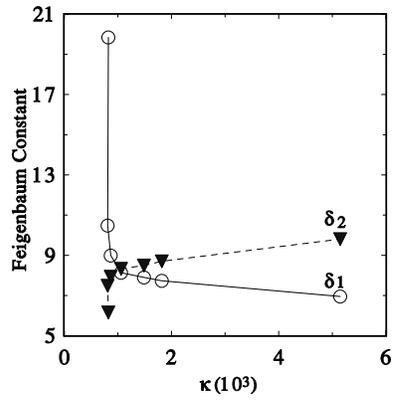,width=8cm,angle=0}}
\caption{The function $\kappa-\delta_{i}$ for period-doubling
bifurcation's in R-H model.}
\end{figure}


\begin{thebibliography}{30}
\bibitem{Glass91}L. Glass,
               Chaos, 
               1 (1991) 13.
\bibitem{He92}D.-R. He {\it et al.}
              Phys. Lett. A
              171 (1992) 61.
\bibitem{Marriot89} C. Marriot and C. Delisle,
              Physica D 36 (1989) 198.
\bibitem{Caiyun98} C. Wu, S. Qu, S. Wu and D.-R. He,
              Chin. Phys. Lett.
              15 (1998) 246.
\bibitem{Xiaoling99} X. Ding, S. Wu, Y. Yin and D.-R. He,
              Chin. Phys. Lett.
              16 (1999) 167.
\bibitem{Feigen78} M.J. Feigenbaum,
               J. Stat. Phys.
                19 (1978) 158.
\bibitem{Hindmarsh82} J. L. Hindmarsh and R. M. Rose,
              Nature(Landon), 296 (1982) 162.

\end{thebibliography}
\end{document}